\documentclass[11pt,twoside]{article}

\usepackage{asp2004}
\usepackage{epsf}
\usepackage{psfig}
\usepackage{lscape}

\def\um{\hbox{$\mu {\rm m}$}}
\def\arcsec{\ifmmode {^{\prime\prime}}\else $^{\prime\prime}$\fi}
\def\arcmin{\ifmmode {^{\prime}}\else $^{\prime}$\fi}
\def\farcs{\ifmmode \rlap.{^{\prime\prime}}\else
    $\rlap.{^{\prime\prime}}$\fi}
\def\arcmper{\ifmmode \rlap.{^{\prime}}\else
    $\rlap.{^{\prime}}$\fi}

\def\etal{{et al.}}
\def\eg{{e.g.,}}

\def\lya{\ifmmode {\rm Ly\alpha}\else{\rm Ly$\alpha$}\fi}
\def\Lya{\ifmmode {\rm Ly\alpha}\else{\rm Ly$\alpha$}\fi}
\def\LFIR{\ifmmode {\rm \,L_{FIR}}\else ${\rm \,L_{FIR}}$\fi}
\def\Lsun{\ifmmode {\rm\,L_\odot}\else ${\rm\,L_\odot}$\fi}
\def\Msun{\ifmmode {\rm\,M_\odot} \else ${\rm\,M_\odot}$\fi}
\def\Msunpyr{\ifmmode {\rm\,M_\odot\,yr^{-1}} \else {${\rm\,M_\odot\,yr^{-1}}$}\fi}
\def\pyr{\ifmmode {\rm\,yr^{-1}} \else {${\rm\,yr^{-1}}$}\fi}
\def\kms{\ifmmode {\rm\,km~s^{-1}} \else ${\rm\,km\,s^{-1}}$\fi}
\def\kmps{\ifmmode {\rm\,km~s^{-1}} \else ${\rm\,km\,s^{-1}}$\fi}
 
\def\ergps{\ifmmode {\rm\,erg\,s^{-1}} \else {${\rm\,erg\,s^{-1}}$}\fi}
\def\ergpspcm{\ifmmode {\rm\,erg\,s^{-1}\,cm^{-2}} \else {${\rm\,erg\,s^{-1}\,cm^{-2}}$}\fi}
\def\surfbr{\ifmmode {\rm\,erg\,s^{-1}\,cm^{-2}\,arcsec^{-2}} \else {${\rm\,erg\,s^{-1}\,cm^{-2}\,arcsec^{-2}}$}\fi}
\def\spose#1{\hbox to 0pt{#1\hss}}
\def\simlt{\mathrel{\spose{\lower 3pt\hbox{$\mathchar"218$}}
     \raise 2.0pt\hbox{$\mathchar"13C$}}}
\def\simgt{\mathrel{\spose{\lower 3pt\hbox{$\mathchar"218$}}
     \raise 2.0pt\hbox{$\mathchar"13E$}}}

\def\ie{{\it i.e. }}

\begin{document}
\title{{\it Spitzer} Observations of High Redshift Radio Galaxies}  
\author{N. Seymour$^1$, D. Stern$^{2,1}$, C. De Breuck$^3$, J. Vernet$^3$, 
  R. Fosbury$^3$, A. Rettura$^{4,3}$, A. Zirm$^5$, B. Rocca-Volmerange$^4$, 
  M. Lacy$^1$, H. Teplitz$^1$, A. Dey$^6$, M. Dickinson$^6$, W. van Breugel$^7$, 
  G. Miley$^8$, H. R\"ottgering$^8$, P. Eisenhardt$^2$, P. McCarthy$^9$, 
  F. De Breuck$^3$, \& L. Vernet$^3$}   

\affil{$^1$Spitzer Science Center, California Institute of Technology,
  Mail Code 220-6, 1200 East California Boulevard, Pasadena, CA 91125 USA,
  $^2$JPL, $^3$ESO, $^4$IAP, $^5$STSCI, $^6$NOAO, $^7$LLNL, $^8$Leiden, 
  $^9$OCIW}

\begin{abstract} 
We present the results of a comprehensive {\it Spitzer} survey of 70 radio
galaxies across $1<z<5.2$. Using IRAC, IRS and MIPS imaging we determine the 
rest-frame AGN contribution to the stellar emission peak at $1.6\mu$m. The 
stellar luminosities are found to be consistent with that of a giant 
elliptical with a stellar mass of $10^{11-12}\,M_\odot$. The mean stellar 
mass remains constant at $\sim10^{11.5}\,M_\odot$ up to $z=3$ indicating 
that the upper end of the mass function is already in place at redshift 3.
The mid-IR luminosities imply bolometric IR luminosities that 
would classify most sources as ULIRGs. The mid-IR to radio luminosity 
generally correlate implying a common origin for these emissions. The ratio 
is higher than that found for lower redshift, \ie $z<1$, radio 
galaxies.
 
\end{abstract}


\section{Introduction}   

Classical radio galaxies are the quintessential type II active galactic 
nuclei (AGN): accreting super-massive black holes that have their continuum 
emission in the UV/optical/NIR absorbed by dust, thus primarily giving them 
the appearance of {\it normal} star-forming galaxies at these wavelengths.
The main evidence that they host super-massive black holes comes
from the high luminosities of their radio 'lobes' that are fed by radio jets 
originating from the host galactic nuclei (Rees 1978). The lobe spatial 
extents (tens of kilo-parsecs) and the luminosities 
($L_{1.4{\rm GHz}}\ge10^{25}$\thinspace WHz$^{-1}$) rule out emission from 
star-formation. To be more precise radio galaxies are Franhoff-Riley type II 
objects with edge-brightened radio lobes. In terms of the orientation 
unification scheme for AGN (Antonucci 1984) radio galaxies are analogous 
to radio loud quasars obscured in the UV/optical/NIR.

Due to their large radio luminosities, radio galaxies were the predominant 
way to probe the distant universe until the advent of ultra-deep optical 
surveys in the last decade. In fact, radio galaxies were the first galaxies 
to be found above redshifts 1, 2, 3 and 4 (\eg\ Stern \&  Spinrad). Since 
their first discovery it has been known that the optical hosts of luminous 
radio sources are primarily giant elliptical (gE and cD) galaxies 
(Matthews \etal\ 1964). In the more distant universe, indirect evidence 
that this association remains intact comes from the detection of normal 
elliptical host galaxies with $r^{1/4}$ law light profiles in $HST$/NICMOS 
observations
of high-redshift radio galaxies (HzRGs) at 1$\simlt$$z$$\simlt$2
(Pentericci et al. 2001; Zirm et al. 2003); the tendency for HzRGs to
reside in moderately rich (proto-)cluster environments (Venemans et
al. 2002; Stern et al. 2003); the spectacular ($>$100\,kpc) luminous
\Lya\ haloes seen around several sources, implying large gas
reservoirs (Reuland et al. 2003; Villar-Mart\'\i n et al. 2003);
sub-mm detections of HzRGs, implying violent star formation activity
up to $\sim$100\,M$_{\odot}$yr$^{-1}$ (Archibald et al. 2001; Reuland
et al. 2004); and a few direct kinematic measurements of HzRGs (Dey \&
Spinrad 1996). The most compelling evidence of this association of
HzRGs with the most massive systems, however, is the tight correlation
of the observed near-infrared Hubble, or $K-z$ diagram for powerful
radio sources (De Breuck et al. 2002, Rocca-Volmerange et
al. 2004): HzRGs form a narrow redshift sequence which traces the
envelope of radio-quiet galaxies and is well-modeled by the evolution
of a stellar population formed at high redshift from a reservoir of
$10^{12}$\,M$_{\odot}$. 
With the more recent
discovery that the stellar bulge and central black hole masses of
galaxies are closely correlated, it is no longer a surprise that the
parent galaxies of the most powerful radio sources occupy the upper
end of the galaxy mass function (Maggorian 1998; Tremaine 2002).

The peak of the stellar emission at $1.6\thinspace\um$ of elliptical 
galaxies has been 
found to be a reasonably robust measure of the stellar mass for old 
passively evolving systems. The mass-to-light ratio at this 
wavelength does not vary greatly for ages $\simgt$ 1Gyr. The {\it Spitzer 
Space Telescope} now allows us to observe this feature
in distant sources in the rest frame. In particular the four bands of the 
IRAC instrument and the IRS $16\thinspace \um$ peak-up imager straddle 
the rest frame $1.6\thinspace \um$ flux density at $1\le$$z$$\le 5$.

\subsection{Sample selection} 

In order to investigate the formation and evolution of the most massive 
galaxies we have performed a {\it Spitzer} survey of 70 HzRGs in GO cycle 1. 
These HzRGs have also been carefully chosen to cover the full range of 
redshifts from $z=1$ to the redshift of the highest known radio galaxy 
($z=5.2$) and two orders of magnitude in radio luminosity, preferentially 
selecting targets with supporting data from $HST$, {\it Chandra} and 
SCUBA/MAMBO. By covering this parameter space, any trends with radio 
luminosity or redshift should be apparent.
The observations consist of photometry in eight bands from the three 
instruments aboard {\it Spitzer}, exercising the full complement of 
imaging capabilities (IRAC: 3.6, 4.5, 5.8 and 8$\,\um$; IRS: $16\,\um$; MIPS:
24, 70 and $160\um$). Due to uncertainty in the ability of 
MIPS to image against the Galactic infrared 
background at the time of submission of GO Cycle 1, we chose only to image 
26/70 HzRGs with MIPS. The other 54 sources have been applied to be 
observed in {\it Spitzer} Cycle 3. The IRS images are only for the 46 objects 
above $z=2$ as below this redshift the $8\,\um$ IRAC channel adequately covers 
the longward side of the $1.6\,\um$ bump. 

The 'SHizRaG' team keep track of this project through a private webpage which 
we intend to make public eventually. Currently a restricted version of the 
webpage is available here:

\noindent
{\tt http://spider.ipac.caltech.edu/staff/seymour/SHizRaGs.html}

\begin{figure}
\begin{minipage}{0.6\linewidth}
\psfig{figure=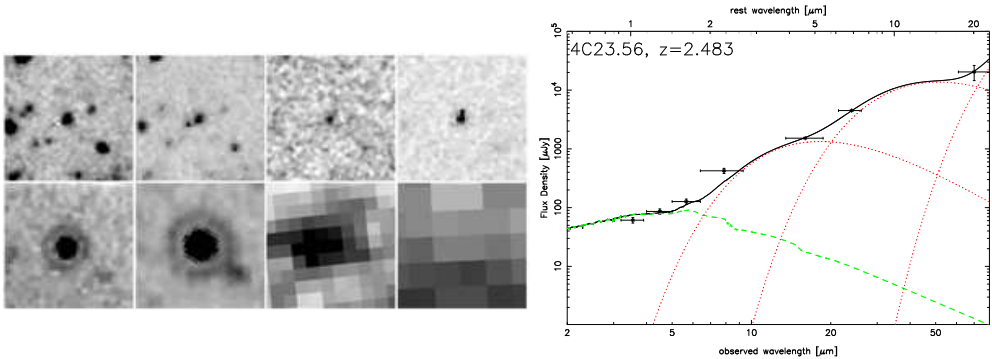,width=7cm,angle=0}
\end{minipage}
\begin{minipage}{0.4\linewidth}
\psfig{figure=4c23.56_pretty.ps,width=6.0cm,angle=-90}
\end{minipage}
\caption{{\bf (Left)} Postage stamp images of 4C23.56 at z=2.48: the 4 
  IRAC bands (top 
  row - wavelength increasing left to right) followed by the IRS $16\,\um$ 
  peak-up image and the MIPS bands (bottom row) showing a clear detection 
  in each waveband out to $70\,\um$ {\bf (Right)} SED fitting of 4C23.56 {\it 
    Spitzer} data using an elliptical galaxy template and black-bodies 
  of various temperatures.}
\end{figure}

\section{Stellar Luminosities and Masses}  

\subsection{SED fitting}

In order to determine the $1.6\,\um$ stellar luminosity, the contribution at 
this wavelength from hot, AGN heated dust needs to be ascertained. We have 
performed this analysis for the sample of 17 MIPS HzRGs which have 
$24\,\um$ detections as well IRAC observations. For the 
modeling presented here we have chosen to use just the {\it Spitzer} data. 
The principle reason is that shorter wavelengths, \ie the rest-frame optical, 
are more likely to have significant contributions from young 
stellar populations, emission lines, and AGN continuum. 

We model the observed 3.6 to $24\,\um$ data using an elliptical stellar 
template and three black-bodies. The elliptical stellar SED is taken from 
the P\'EGASE (Fioc \& Rocca-Volmerange 1997) software and corresponds 
to a passively evolving $10^{12}\,M_\odot$ stellar population 
formed at $z=10$; \ie for each redshift a template of the correct age was 
chosen although the SED does not evolve significantly after 1\,Gyr (\ie 
$z\sim4.4$). The three black-bodies where chosen to be at 60\,K (analogous to 
the temperature of cold dust found from sub-mm observations), 250\,K and 
600-1200\,K. This last value was allowed to vary and the best temperature from 
the fitting applied. Full details of this modeling are presented in Seymour 
\etal\ ({\it in prep.}). 

Figure 1 illustrates the {\it Spitzer} observations and the SED modeling for 
one representative source, 4C23.56 at $z=2.48$. The best fit hot dust 
temperature is 750\,K and one can see that the 1.6$\,\um$ stellar peak has 
considerable AGN contamination. A more detailed analysis of the broad-band, 
X-ray to radio SED of 4C23.56 is presented in De Breuck \etal\ ({\it in 
  prep.}).

The other 53 HzRGs with no MIPS detections can at least provide upper limits 
to the stellar luminosities and, in some cases only the upper limits come 
from detections only in IRAC 
channels 1 and 2. So we fit a maximum elliptical template SED to the IRAC 
data. In some case the elliptical template fits quite well to IRAC channels 
1-3 and in others the SED is steeply rising at the longer wavelengths 
and the fit is restricted only by channel 1. In the former case the we 
calculate a 'nominal' stellar mass from the fit, but in the latter we may 
only derive upper limits.

\begin{figure}
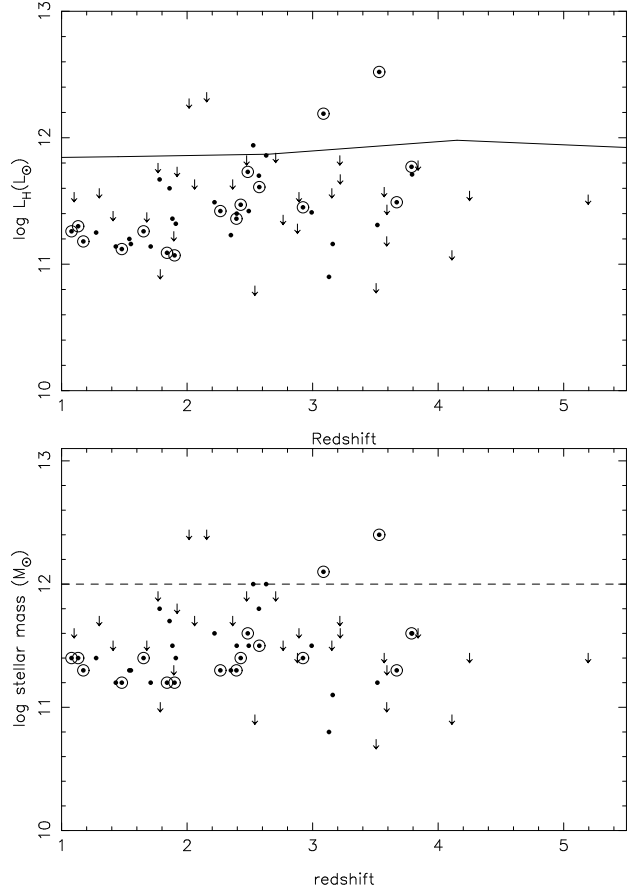

  \centering
  \begin{minipage}{1.0\linewidth}
    \psfig{figure=loghz.ps,width=8.2cm,angle=-90}
  \end{minipage}
  \begin{minipage}{1.0\linewidth}
    \psfig{figure=logmz.ps,width=8.2cm,angle=-90}
  \end{minipage}
  \caption{{\bf (Top)} HzRG Stellar luminosities plotted against redshift of 
    each HzRG. The solid line indicates a of $10^{12}\,M_\odot$ elliptical 
    galaxy. {\bf (Bottom)} Stellar masses from the SED fitting plotted 
    against redshift. Most stellar luminosities indicate with stellar masses 
    of $10^{11}-10^{12}\,M_\odot$. Stellar luminosities and masses derived 
    from sources with MIPS $24\,\um$ detections are circled, the dots with 
    out circles are 'nominal' masses from just IRAC data. Downward arrows 
    indicate upper limits.}
\end{figure}

\subsection{Results}

The derived rest-frame, AGN-subtracted, 1.6$\,\um$ stellar luminosities are 
shown against redshift in Fig. 3 (top) for all our HzRGs. Also laid on the 
plot by a solid line is the stellar luminosity of a $10^{12}\,M_\odot$ 
elliptical galaxy. The stellar luminosities imply stellar masses in the range
$10^{11}-10^{12}\,M_\odot$ with a mean mass of $\sim10^{11.5}\,M_\odot$ (Fig. 
3 bottom). This mean mass remains consistent out to $z=3$ (beyond which 
the parameter space becomes less well sampled) suggesting that the upper 
end of the mass function is already in place by at least $z=3$.

\section{Mid-IR luminosities}

Figure 3 shows the rest 6.75$\,\um$ luminosity against the rest 3\,GHz 
luminosity for the 18 radio galaxies with MIPS observations. The wavelength 
of 6.75$\,\um$ was chosen as a fiducial mid-IR wavelength as it is the mean 
rest wavelength of the observed MIPS 24$\,\um$ band for our sample
and also the wavelength of the LW2 filter from ISOCAM, allowing a direct 
comparison of derived luminosities. It is clear to first approximation that 
the two luminosities correlate and by implication have a common origin. This
correlation makes sense if the radio luminosity comes from lobes induced by a 
jet from the AGN and the mid-IR comes from hot AGN-heated dust.

The mid-IR luminosities are also all greater than $10^{11}L_\odot$ 
implying bolometric luminosities on ULIRG scales assuming local 
relations hold for these objects. These radio galaxies tend to have a higher 
mean mid-IR to radio luminosity ratio then those selected at lower redshift, 
\eg\ Ogle \etal\ (2006) finds that $z<1$ radio galaxies tend to have
$L_{\rm mid-IR}/L_{\rm radio}\sim10-100$. 

\begin{figure}
\centering
\begin{minipage}{1.0\linewidth}
    \psfig{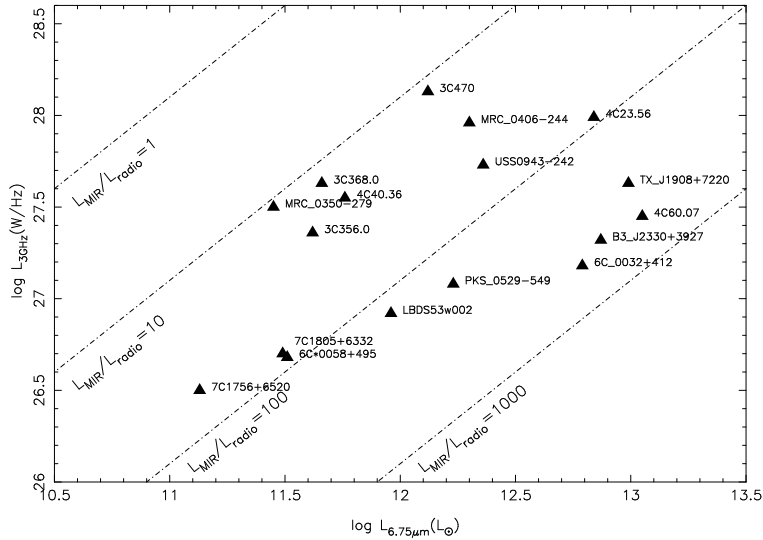}
\end{minipage}
\caption{Rest mid-IR luminosity against rest radio luminosity which 
  approximately correlate implying a common origin. This correlation may be 
  explained by the AGN producing the radio jets and also heating the hot 
  dust radiating in the mid-IR. The mid-IR/radio ratio increases at higher 
  redshifts compared to low redshift samples (\eg\ Ogle \etal\ 2006) most 
  likely due to the hotter temperatures of the AGN-heated dust.}
\end{figure}

\section{Conclusions and future work}

We have presented a stellar-luminosity/redshift relation of HzRGs, a 
more physical representation of the $K-z$ diagram. This distribution seems to 
confirm the long held paradigm that radio galaxies are hosted by massive 
ellipticals out to high redshifts, and that the most massive galaxies are 
already in place
by redshift 4 and possibly earlier. We also observe a correlation between the 
infrared and radio luminosities which is unsurprising if they are both 
fueled by the AGN.

Current on-going work includes mm/sub-mm observations with SCUBA, MAMBO and 
the CSO to constrain the cold dust component at longer wavelengths and hence 
estimate the mass of this cold dust. Over-densities of sources around radio 
galaxies are being investigated to look for evidence of cluster formation 
(Zirm \etal\, {\it in prep.}). 
Over half the sources with $24\,\um$ images have over-densities of factors of 
2-5 greater than that expected from 24$\,\um$ source counts. These radio 
galaxies mainly lie at $1.5<z<2.5$ where the strong $6-8\,\um$ PAH feature 
passes though the 24$\,\um$ MIPS band, enhancing the 24$\,\um$ flux density.

%
%
%
%
%
%
%
%


\acknowledgements 
We thank the LOC for organising a great conference and were particularly 
impressed by the nifty design of the webpage. We also liked Harry's 
conference haircut.


\end{document}